\documentclass[aps,twocolumn,showpacs]{revtex4}
\usepackage{epsfig}
\usepackage{amsbsy}
\usepackage{amsmath}
\usepackage{amsfonts}
\usepackage{graphicx}
\usepackage{latexsym}
\usepackage{graphics}
\usepackage{caption2}

\newcommand{\be}{\begin{equation}}
\newcommand{\ee}{\end{equation}}

\begin{document}

\title{Matrix realignment and partial transpose approach to entangling power
of quantum evolutions}
\author{Zhihao Ma$^{1,2,3}$ and Xiaoguang Wang$^{1}$}
\affiliation{1. Zhejiang Institute of Modern Physics, Department of Physics, Zhejiang
University, Hanzhou 310027, China.}
\affiliation{2. Department of Mathematics, Shanghai Jiaotong University, Shanghai,
200240, China}
\affiliation{3. Department of Mathematics, Zhejiang University, Hanzhou 310027, China.}
\date{\today}

\begin{abstract}
Based on the matrix realignment and partial transpose, we develop an
approach to entangling power and operator entanglement of quantum unitary
operators. We demonstrate efficiency of the approach by studying several
unitary operators on qudits, and indicate that these two matrix
rearrangements are not only powerful for studying separability problem of
quantum states, but also useful in studying entangling capabilities of
quantum operators.
\end{abstract}

\pacs{03.67.-a,03.65.Ud}
\maketitle

Given a unitary operator, in the context of quantum information~\cite{Nie00}%
, one may ask how much entanglement capability does the operator have. The
entangling unitary operator can be considered as a resource for quantum
information processing, and it becomes important to quantitatively describe
unitary operators. Recently, there are increasing interests in the
entanglement capabilities of quantum evolutions and Hamiltonians \cite{Zan00}%
-\cite{Kra01}. The entangling power based on the linear entropy~\cite{Zan00}
is a valuable, and relatively easy to calculate, measure of the entanglement
capability of an operator. The entangling power for two qudits can be
expressed in terms of operator entanglement~\cite{Zan01,Wan02} (also called
Schmidt strength~\cite{Nie02}). Both entangling power and operator
entanglement have been applied to the study of quantum chaotic systems ~\cite%
{Chao1,Baker1,Baker2,Baker3}. Moreover, the concept of entangling power has
been extended to the case with ancillas~\cite{Wang2}, the case of
entanglement-changing power~\cite{Guo}, and the case of disentangling power
\cite{Sudbery}.

Let us start by introducing some basics of entanglement of quantum states,
the operator entanglement, and the entangling power. For a two-qudit pure
state $|\Psi \rangle \in \mathcal{H}_{d}\otimes \mathcal{H}_{d},$ one can
quantify entanglement by using the linear entropy
\begin{equation}
E(|\Psi \rangle ):=1-\text{Tr}\rho _{1}^{2},  \label{e2}
\end{equation}%
where $\rho _{1}=\text{Tr}_{2}(|\Psi \rangle \langle \Psi |)$ is the reduced
density matrix. The linear entropy satisfy the inequalities $\quad 0\leq
E(|\Psi \rangle )\leq 1-1/d$, where the lower (upper) bound is reached if
and only if $|\Psi \rangle $ is a product state (maximally entangled state).

In the orthogonal basis $\{|1\rangle ,...,|d\rangle \}$, state $|\Psi
\rangle $ is written as
\begin{equation}
|\Psi \rangle =\sum_{i,j=1}^{d}A_{ij}|i\rangle \otimes |j\rangle ,
\label{state}
\end{equation}%
where $A_{ij}$ are the coefficients, and $A$ can be considered as a matrix.
After direct calcualtions, one find that the reduced density matrix $\rho
_{1}=AA^{\dagger }. $ Substituting it to Eq. (\ref{e2}) leads to another
expression of the linear entropy
\begin{equation}
E(|\Psi \rangle )=1-\text{Tr}\left( AA^{\dagger }AA^{\dagger }\right),
\label{entropy1}
\end{equation}%
which will be used for later discussions.

An operator can increase entanglement of a state, but an operator can also
be considered to be entangled because operators themselves inhabit a Hilbert
space. The entanglement of quantum operators is introduced \cite{Zan01} by
noting that the linear operators over $\mathcal{H}_{d}$ span a $d^{2}$%
-dimensional Hilbert space with the scalar product between two operators $X$
and $Y$ given by the Hilbert-Schmidt product $\langle X,Y\rangle :=\text{Tr}%
(X^{\dagger }Y)$, and $||X||_{\mathrm{HS}}:=\sqrt{\text{Tr}(X^{\dagger }X)}$%
. We denote this $d^{2}$-dimensional Hilbert space as $\mathcal{H}_{d^{2}}^{%
\text{HS}}$. Thus, the operator acting on $\mathcal{H}_{d}\otimes \mathcal{H}%
_{d}$ is a state in the composite Hilbert space $\mathcal{H}_{d^{2}}^{\text{%
HS}}\otimes \mathcal{H}_{d^{2}}^{\text{HS}}$, and the entanglement of an
operator $X$ is well-defined \cite{Zan01}.

The entangling power quantifies the entanglement capability of a unitary
operator $U$. It is defined as~\cite{Zan00}
\begin{equation}
e_{\text{p}}(U):=\overline{E(U|\psi _{1}\rangle \otimes |\psi _{2}\rangle )}.
\end{equation}%
It tells us how much entanglement the operator produces, on average, when
acting on product states. After a suitable average over initial product
states, one find~\cite{Zan00}
\begin{equation}
e_{\text{p}}(U)=\left( \frac{d}{d+1}\right)^{2}\left[
E(U)+E(US_{12})-E(S_{12})\right] .  \label{ep}
\end{equation}%
Thus, the entangling power defined on $d\times d$ systems can be expressed
in terms of the entanglement of three operators, $U$, $US_{12}$, and $S_{12}$%
. Here, $S_{12}$ is the swappig operator. Therefore, by studying the
entanglement of these three operators we can obtain the entangling power of $%
U$.

Next, we give our approach, and first consider the operator entanglement of
a unitary operator. A unitary operator can be written as
\begin{eqnarray}
U &=&\sum_{ijkl}\langle ij|U|kl\rangle |ij\rangle \langle kl| \\
&=&\sum_{ijkl}U_{ij,kl}|i\rangle \langle k|\otimes |j\rangle \langle l| \\
&=&\sum_{ijkl}U_{ij,kl}e_{ik}\otimes e_{jl},
\end{eqnarray}
where $e_{ik}$ are orthogonal basis in the space $\mathcal{H}_{d^{2}}^{\text{%
HS}},$ and they can be considered as states. Now, we define a new matrix $%
U^{R}$ as
\begin{equation}
\left( U^{R}\right) _{ij,kl}=U_{ik,jl}.
\end{equation}%
The matrix can be obtained by realigment of matrix $U.$Using this realigned
matrix, one can express normalized unitary operator $\tilde{U}$ as
\begin{equation}
\tilde{U}=\frac{U}{d}=\frac{1}{d}\sum_{ijkl}\left( U^{R}\right)
_{ik,jl}e_{ik}\otimes e_{jl},  \label{state2}
\end{equation}%
Comparing Eqs.(\ref{state}) and (\ref{state2}), and using Eq. (\ref{entropy1}%
), one obtain the operator entanglement of $U$

\begin{equation}
E(U)=1-\frac{1}{d^{4}}\text{Tr}(U^{R}\left( U^{R}\right) ^{\dagger
}U^{R}\left( U^{R}\right) ^{\dagger }).  \label{eq1}
\end{equation}%
We see that the operator entanglement is determined by the naturally
appeared realigned matrix. The realigned matrix is easy to obtain from the
original unitary matrix, and thus our approach is very efficient to study
operator entanglement.

It is more interesting to see that this matrix realigment is the same as
density matrix realigment when studying the separability problem of quantum
mixed state \cite{Crossnorm}. The realignment criteria (also called cross
norm criteria) is very strong to detect many bound entangled states. We see
here that the same matrix realignment approch is very effective in studying
operator entanglement.

There is another matrix rearrangement, called partial transpose~\cite{PT}. A
partial transpose with respect to the first system $U^{T_{1}}$ is defined as
\begin{equation}
\left( U^{T_{1}}\right) _{ij,kl}=U_{kj,il}.
\end{equation}%
It is well-known that the partial transposed method is very useful in
studying entanglement of quantum mixed states. Was it useful in studying
operator entangling properties? We will see that indeed it is.

The entangling power is determined by three operator entanglement $%
E_{l}(U),E_{l}(S_{12}),$ and $E_{l}(S_{12}U)$. The first two can be
determined by the realigment method, and the last one is of course can be
determined by the same method, but with extra effort to make matrix
multiplication $S_{12}U.$ In fact, we have \cite{Fan}
\begin{equation*}
S_{12}\left( S_{12}U\right) ^{R}=U^{T_{1}}.
\end{equation*}%
Using the above property and applying Eq.(\ref{eq1}) to $S_{12}U,$ we obtain
\begin{equation}
E(S_{12}U)=1-\frac{1}{d^{4}}\text{Tr}(U^{T_{1}}\left( U^{T_{1}}\right)
^{\dagger }U^{T_{1}}\left( U^{T_{1}}\right) ^{\dagger })  \label{eq4}
\end{equation}%
Therefore, the operator entanglement of $S_{12}U$ can be written in terms of
partial transposed unitary matrix $U^{T_{1}}.$

From Eqs. (\ref{ep}), (\ref{eq1}), and (\ref{eq4}), we know that the
entangling power can be determined by the matrix realignment and the partial
transpose
\begin{align}
&e_{\text{p}}(U)=\left( \frac{d}{d+1}\right) ^{2}\left[ 2-E(S_{12})\right]
\notag \\
&-\frac{1}{(d+1)^{2}d^{2}}\text{Tr}(\left[ U^{R}\left( U^{R}\right)
^{\dagger }\right] ^{2}+\left[ U^{T_{1}}\left( U^{T_{1}}\right) ^{\dagger }%
\right] ^{2}).
\end{align}%
Both these matrix manipulations are powerful in the context of separability
of quantum states. Here, we find they are also powerful in studying operator
entanglement and entangling power in quantum information theory.

To illustrate the efficiency of the approach, let us consider several
examples.

\textit{Example 1}: The swap operator $S_{12}.$ It can be written as
\begin{equation}
S_{12}=\sum\limits_{i,j=1}^{d}|ij\rangle \langle ji|.
\end{equation}%
It is easy to see that
\begin{equation*}
S_{12}^{R}=S_{12},S_{12}^{\dagger }=S_{12},S_{12}^{2}=I.
\end{equation*}%
The swap operator is invariant under the matrix realigment. Then, from Eq.(%
\ref{eq1}), the linear entropy of the swap operator is given by
\begin{equation}
E_{l}(S_{12})=1-\frac{1}{d^{4}}\text{Tr}(S_{12}^{4})=1-\frac{1}{d^{2}}.
\end{equation}%
From Eq.(\ref{ep}), evidently the entanging power of the swap operator is
zero.

\textit{Example 2}: The unitary operator $V$ generated by the swap
\begin{equation}
\quad V=\exp (-itS_{12})=\cos (t)I-i\sin (t)S_{12}.
\end{equation}%
It is straightfoward to check the following identities
\begin{equation}
I^{T_{1}}=I,I^{R}=dP_{+},S_{12}^{T_{1}}=dP_{+},S_{12}^{R}=S,
\end{equation}%
where projector
\begin{equation}
P_{+}=|\Psi _{+}\rangle \langle \Psi _{+}|,|\Psi _{+}\rangle =\frac{1}{\sqrt{%
d}}\sum_{i=1}^{d}|i\rangle \otimes |i\rangle .
\end{equation}%
From the above identies, we obtain
\begin{eqnarray}
\quad V^{R}(t) &=&\cos (t)dP_{+}-i\sin (t)S_{12},  \notag \\
V^{T_{1}}(t) &=&\cos (t)I-i\sin (t)dP_{+}.
\end{eqnarray}%
Then, we find
\begin{eqnarray}
\quad V^{R}(t)\left[ V^{R}(t)\right] ^{\dag } &=&\cos ^{2}(t)d^{2}P_{+}+\sin
^{2}(t)I,  \notag \\
V^{T_{1}}(t)\left[ V^{T_{1}}(t)\right] ^{\dag } &=&\cos ^{2}(t)I+\sin
^{2}(t)d^{2}P_{+}.
\end{eqnarray}%
From the above two equations and Eqs. (\ref{eq1}), and (\ref{eq4}), we find
linear entropies
\begin{eqnarray}
E(V) &=&\left( 1-\frac{1}{d^{2}}\right) (1-\cos ^{4}t)  \label{eee} \\
E(VS_{12}) &=&\left( 1-\frac{1}{d^{2}}\right) (1-\sin ^{4}t)  \label{eeee}
\end{eqnarray}%
Substuting Eqs.(\ref{eee})and (\ref{eeee}) into (\ref{ep}) leads to the
expression of the entangling power
\begin{equation}
e_{\text{p}}=\frac{d^{2}-1}{2(d+1)^{2}}\sin ^{2}(2t).  \label{eeeee}
\end{equation}

From Eqs.(\ref{eee}) and (\ref{eeeee}), we see that the maximal value of the
operator entanglement occurs at $t=\frac{\pi }{2},$ however, at this point
the entangling is zero. This point corresonds to the swap. The maximal
entangling power occurs at $t=\frac{\pi }{4},$ which corresponds to the $%
\sqrt{swap}$ gate, the square of which is just the swap gate. Thus, the $%
\sqrt{swap}$ gate can be used as an important gate for quantum computing not
only in qubit systems, but also in qudit systems. Quantiatively, the
operator entanglement and entangling power of the $\sqrt{swap}$ gate is
given by
\begin{equation}
E(V)=\frac{3}{4}\left( 1-\frac{1}{d^{2}}\right) ,e_{\text{p}}=\frac{d^{2}-1}{%
2\frac{\frac{{}}{{}}}{{}}(d+1)^{2}}.
\end{equation}%
respectively.

\noindent \textit{Example 3}: \ A general two-qudit controlled-$U$ gate is
given by
\begin{equation}
C_{U}:=\sum_{n=1}^{d}|n\rangle \langle n|\otimes U_{n},  \label{cu}
\end{equation}%
The controlled-$U$ gate implements the unitary operator $U_{n}$ on the
second system if and only if the first system is in the state $|n\rangle $.
For the controlled-$U$ operation, it was found that \cite{Wang2}
\begin{equation}
e_{\text{p}}(C_{U})=\left( \frac{d}{d+1}\right) ^{2}E(C_{U}),  \label{ff}
\end{equation}%
Let us prove this via our approach. From Eq.(\ref{ep}), to prove the above
identity is equivalent to prove that
\begin{equation}
E(C_{U}S_{12})=E(S_{12}).  \label{equ}
\end{equation}%
In fact, we have a more general result that \textit{if the partial transpose
of a unitary operator }$U$\textit{\ is still an unitary operator, then, }$%
E(US_{12})=1-1/d^{2}=E(S_{12}).$This result immediately follows from Eq. (%
\ref{eq4}). For our operator $C_{U},$ from the definition, it is not
diffulcult to see that it is invariant under the partial transpose with
respect to the first system. Of course, $C_{U}$ is unitary, and then Eq.(\ref%
{equ}) holds. In this case, the entangling power is proportional to the
operator entanglement of the controlled-$U$ gate. It is easier to obtain Eq.(%
\ref{equ}) via our approach.

In conclusion, we have developed an efficient way for studying entangling
power and operator entanglement. One only needs to obtain the realigned
unitary operator and partially transposed operator to determine the
entangling power. Once we have analytical expression for a unitary matrix,
then analytical expressions for entangling power and operator entanglement
can be obtained. If we cannot have the analytical expression, it is very
convenient to make the matrix rearrangements numerically, and then
entangling power and operator entanglement can be quickly computed.

The matrix realignment and partial transpose play very important roles in
the theory of separability of quantum mixed states, and we see here that
they naturally appears in the study of entanglement capabilities of quantum
evolutions. The approach developed here can be applied to investigate
entanglement capabilities in many physical systems such as quantum chaotic
systems.

\acknowledgements This work was supported by CNSF under grant no. 10405019,
Specialized Research Fund for the Doctoral Program of Higher Education
(SRFDP) under grant No.20050335087, and The Project-sponsored by SRF for
ROCS, SEM.

\end{document}